\documentclass[prd,a4paper,showpacs,byrevtex,preprint]{revtex4-1}
\usepackage{graphicx}

\usepackage{amsmath}
\usepackage{array}
\usepackage{bm}
\usepackage{amssymb}
\usepackage{amsfonts}

\begin{document}

\title{Anyons in quantum mechanics with a minimal length}

\author{Fabien \surname{Buisseret}}
\email[E-mail: ]{fabien.buisseret@umons.ac.be}
\affiliation{Haute \' Ecole Louvain en Hainaut (HELHa), Chauss\'ee de Binche 159, 7000 Mons, Belgium} 
\affiliation{Service de Physique Nucl\'{e}aire et Subnucl\'eaire,
Universit\'{e} de Mons -- UMONS,
Place du Parc 20, 7000 Mons, Belgium}


\date{\today}

\begin{abstract}
The existence of anyons, \textit{i.e.} quantum states with an arbitrary spin, is a generic feature of standard quantum mechanics in $(2+1)-$dimensional Minkowski spacetime. Here it is shown that relativistic anyons may exist also in quantum theories where  a minimal length is present. The interplay between minimal length and  arbitrary spin effects are discussed.
\end{abstract}

\pacs{02.40.Gh; 03.65.Ge; 05.30.Pr}


\maketitle

\section{Introduction}

The existence of a minimal observable length in Nature is suggested by string theory and quantum gravity, see \textit{e.g.}~\cite{s1}. An economical way of introducing such a minimal length at the level of nonrelativistic quantum mechanics is to use a suitably modified Heisenberg algebra~\cite{k0a,k0b,k0c}. This proposition has been extended to covariant quantum mechanics in \cite{Quesne2006} and further investigated in many relevant cases including the Klein-Gordon \cite{Moayedi:2010vp} and Dirac \cite{Moayedi:2011ur} equations. 

Although the study of quantum theories characterized by a minimal length has become an active field of research, some comments remain to be done about the case of a $(2+1)-$dimensional spacetime with a minimal length. Indeed, it is well-known that spin is no longer quantized in $2+1$ dimensions, allowing the appearance of anyons, \textit{i.e.} paricles with arbitrary spin. To our knowledge, the question of the existence (or not) of anyons in a $(2+1)-$dimensional spacetime with a minimal length has never been addressed.  It is the topic of the present paper.

\section{Poincar\'e algebra in $2+1$ dimensions}\label{poinc}
\subsection{Generalities}
The Poincar\'e algebra in $2+1$ dimensions can be written under the form \cite{Barut:1965}
\begin{equation}\label{alg0}
[J^\mu ,J^\nu]= - i \varepsilon^{\mu\nu\rho} J_\rho\;, \quad 
[J^\mu, P^\nu]= -i \varepsilon^{\mu\nu\rho} P_\rho;, \quad 
[P^\mu ,P^\nu]=0\; ,
\end{equation}
where $P^\mu$ are the translation generators and where 
\begin{equation}
J^\mu=\frac{1}{2}\epsilon^{\mu\nu\rho}L_{\nu\rho},
\end{equation}
$L_{\mu\nu}$ being the Lorentz generators. The convention $\epsilon^{012}=1$ is used, as well as the Minkowski metric $\eta=$diag$(+,-,-)$. The two Casimir operators associated to (\ref{alg0}) read, for massive representations,
\begin{equation}\label{casimir}
M^2={\bm P}^2, \quad s=\frac{{\bm P}\cdot {\bm J}}{M}.
\end{equation}
They represent the squared mass and the spin of a given state respectively, while the notation ${\bm X}=(X^0,X^1,X^2)$ for a $3-$vector.

It is useful to note that, in the rest frame of a given state, $s$ reduces to $J^0=\pm L_{12}$, that is the generator of spatial rotations. One recovers then a maybe more intuitive definition of spin as the phase factor associated to rotations. 

The unitary irreducible representations of the Lorentz algebra in $2+1$ dimensions -- the first commutator of  (\ref{alg0}) -- have been built in \cite{Barut:1965}, while the unitary irreducible representations of the full Poincar\'e algebra have been obtained in \cite{Binegar:1981gv}. The conclusions shared by both studies are identical in what concerns the spin of the representations: It can be an arbitrary real number. This is a peculiar feature of quantum mechanics in $2+1$ dimensions and strongly differs from the $(3+1)-$ dimensional case, in which the spin of a state can only be integer of half-integer.  The properties of such particles with arbitrary spin, called anyons, have been studied in a considerable amount of works. The interested reader may find relevant informations in the pioneering works \cite{Leinaas} and  \cite{Wilczek,Wil2}, as well as in the reviews \cite{Fort,Khare}. 

It is known that finite-dimensional representations of the Lorentz algebra are non-unitary in $2+1$ dimensions. However, this does not imply that bosons and fermions are fobidden. It is indeed worth mentioning that bosonic  and fermionic states can be built in $2+1$ dimensions from the tensor product of non-unitary infinite-dimensional representations of the Lorentz algebra. That subtle issue has been studied in \cite{Boson}, to which we refer the interested reader.  

\subsection{Minimal length representation of the Poincar\'e algebra}
It has been shown in \cite{Quesne2006} that, for a $(D+1)-$dimensional spacetime  with metric $\eta=$diag$(+,-,\dots,-)$, the Poincar\'e algebra is represented by the generators
\begin{equation}\label{ML}
P_\mu=(1-\beta \hat {\bm P}^2)^{-1} \hat P_\mu,\quad L_{\mu\nu}=(1-\beta \hat {\bm P}^2)^{-1}(\hat P_\nu \hat X_\mu -\hat P_\mu \hat X_\nu )
\end{equation}
provided that 
\begin{eqnarray}\label{alg1}
\left[ \hat X^\mu,\hat P^\nu\right] &=&-i\left[ (1-\beta \hat{\bm P}^2)\eta^{\mu\nu}-\beta' \hat P^\mu \hat P^\nu\right] ,\nonumber \\
\left[  \hat X^\mu, \hat X^\nu\right] &=& - i \left[ (2\beta-\beta')-(2\beta+\beta')\beta \hat{\bm P}^2 \right] \ L^{\mu\nu}, \\
\left[ \hat P^\mu,\hat P^\nu\right] &=& 0 \nonumber .
\end{eqnarray}
where it is assumed that $\beta$, $\beta'\in \mathbb{R}^+$ .
The modified Heisenberg algebra (\ref{alg1}) leads to a minimal uncertainty on the measurement of a position that can be computed to be, assuming isotropicity \cite{Quesne2006},
\begin{equation}
\Delta X_{{\rm min}}=\sqrt{(D\beta+\beta')(1-\beta \langle (P^0)^2\rangle)} ,
\end{equation} 
and is usually referred to as ``minimal length".

\section{Relativistic anyons with a minimal length}\label{sec2}

It follows from the general results recalled in the previous section that, in $(2+1)$ dimensions, quantum theories which are both Poincar\'e invariant and formulated in a spacetime with minimal length are allowed. As a consequence, anyons may exist when a minimal length is present too. Wave equations describing an anyon $\lvert \psi\rangle$ have been proposed in \cite{cortes} in a form that is convenient  regarding to our framework. They read
\begin{equation}\label{aneq}
V_\mu \lvert\psi\rangle=0, \quad V_\mu=s P_\mu-i \epsilon_{\mu\nu\rho} P^\nu J^\rho+M J_\mu .
\end{equation}
They are valid whatever the representation chosen for $P^\mu$ and $J^\mu$, so they can be used with the minimal length representation (\ref{ML}), (\ref{alg1}). An extensive discussions of the solutions of (\ref{aneq}) can be found in \cite{cortes}.

In view of explicit computations, it is convenient to note that the algebra (\ref{alg1}) can be represented by the operators \cite{Quesne2006}
\begin{eqnarray}\label{rep}
\hat P^\mu&=& p^\mu , \nonumber \\
\hat X^\mu &=& (1-\beta {\bm p}^2) x^\mu-\beta' p^\mu {\bm p}\cdot {\bm x}+ i \gamma p^\mu , 
\end{eqnarray}
where $x_\mu=-i \frac{\partial}{\partial p^\mu} $ and ${\bm p}$ are standard position and momentum operators, \textit{i.e.} $\left[ x^\mu,p^\nu\right] =-i \eta^{\mu\nu}$. A straightforward computation shows that, using this representation, the Lorentz generators take a more familiar form
\begin{equation}\label{Lmn}
L_{\mu\nu}=i(p_\mu \frac{\partial}{\partial p^\nu}-p_\nu \frac{\partial}{\partial p^\mu}).
\end{equation} 
 Using the representation (\ref{rep}) in polar coordinates, namely $p^\mu=(p^0,p,\theta_p)$, one quickly shows that 
\begin{equation}\label{j0}
J^0=L_{12}=i\frac{\partial}{\partial_{\theta_p}},
\end{equation}
recovering a standard form for $J^0$ even if a modified Heisenberg algebra is considered. 

A peculiarity of algebra (\ref{alg1}) is that $\left[ \hat X^\mu,\hat X^\nu \right] \propto L^{\mu\nu}$. Hence, the noncommutativity of spatial coordinates is such that $\left[ \hat X^1,\hat X^2 \right] \propto J^0$.  Let us indeed consider an anyonic state of squared mass $M^2$ and spin $s$ in its rest frame and denote $\lvert \psi; M^2,s,j\rangle$ such a state. Then one is led to the following uncertainty relation on the spatial coordinates:
\begin{equation}\label{nc}
\Delta X^1\, \Delta X^2 \geq \frac{1}{2} \lvert (2\beta-\beta')-(2\beta+\beta')\beta M^2 \rvert \, s. 
\end{equation}
Anyons with larger spins then ``feel" a larger spatial noncommutativity. Note however that, even for small values of $s$, one has the lower bound $\Delta X^1 \Delta X^2 \geq \Delta X_{{\rm min}}^2$ implied by the modified Heisenberg algebra. 

\section{Nonrelativistic limit}\label{secnr}

The modified Heisenberg algebra initially proposed by Kempf in a nonrelativistic version \cite{k0a} can be recovered from algebra (\ref{alg1}) and from (\ref{ML}) by neglecting the terms in $\beta (P^0)^2$, \textit{i.e.} by assuming that the typical energy of the system under study is negligible compared to $1/\sqrt\beta$, and by only considering the spatial commutators. The only nonvanishing commutators in the case of two spatial dimensions are then
\begin{eqnarray}\label{alg2}
\left[ \hat X^i,\hat P^j\right] &=&i\left[ (1+\beta  \hat P^2)\delta^{ij}+\beta' \hat P^i \hat P^j\right] ,\nonumber \\
\left[  \hat X^1, \hat X^2\right] &=& -i \left[ (2\beta-\beta')+(2\beta+\beta')\beta \hat P^2 \right]  J^0 ,
\end{eqnarray}
where $\hat P^2=(P^1)^2+(P^2)^2$. The momentum representation (\ref{rep})  has the nonrelativistic counterpart
\begin{eqnarray}\label{rep2}
\hat P^j&=& p^j , \nonumber \\
\hat X^j &=& i(1+\beta  p^2) \frac{\partial}{\partial_{p^j}}+i \beta' p^j \vec p\cdot \vec \nabla_p+ i \gamma p^j .
\end{eqnarray}

It can be check from the above representation that any Hamiltonian of the form $T(\hat P^2)+V(\hat X^2)$, being spherically symmetric in momentum representation, will commute with $J^0$, according to (\ref{j0}). Moreover, in the rest frame of the eigenstate $\lvert \chi \rangle$ under study, $J^0$ is identified with the spin; its eigenvalue $s$ can then be arbitrary. It follows 
\begin{equation}\label{ang}
\langle \vec p\vert  \chi\rangle= {\rm e}^{i s \theta_p} \psi(p),
\end{equation}
which fixes the angular dependence of the state. This last relation also shows that, if $\vec{ \hat P}$ and $\vec{ \hat X}$ are the relative momentum and position of a two-body system made of two identical bodies, the wave function acquires a phase factor ${\rm e}^{i s\pi}$ under permutation of the two bodies. This exotic phase factor is the signature of a braid statistics, in agreement with the generalized spin-statistics theorem \cite{mund}. 

As an illustration, let us consider the Hamiltonian $\frac{\hat P^2}{2\mu}+\frac{1}{2}\mu \omega^2 \hat X^2$, whose energy spectrum is analytically known in $D$ spatial dimensions \cite{minic}. When $D=2$, the spin is arbitrary and, using the results of this last reference, the energy spectrum is given by 
\begin{eqnarray}
\frac{E}{\omega}&=&(2n+\vert s\vert +1)\sqrt{1+\left[\beta^2 s^2 + \frac{(2\beta+\beta')^2}{4}\right]\mu^2 \omega^2 }\nonumber \\
&&+ \left[ (\beta+\beta') (2n+\vert s\vert +1)^2 + (\beta-\beta') (s^2+1)+\beta'\right] \frac{\mu\omega}{2},
\end{eqnarray}
which is a generalization of the anyonic harmonic oscillator, that can be recovered with $\beta=\beta'=0$, see \textit{e.g.} \cite{Wil0,Khare}. The main effect of the modified Heisenberg algebra is to break to well-known degeneracy in $2n+\vert s\vert +1$. 

It is worth pointing out that there exists a deformation of the Heisenberg algebra which is inequivalent to (\ref{alg2}) but which can also be related to anyons. It reads 
\begin{eqnarray*}
[ a^-,a^+]&=&1 +  \nu K, \\
\{K,a^\pm \}&=&0,\quad K^2=1,
\end{eqnarray*}
where the real parameter $\nu$ and the kleinian $K$ are responsible for the deformation. $a^\pm$ are creation and annihilation operators. As shown in \cite{Plyu}, this algebra can be used to describe in an elegant way relativistic quantum states with arbitrary spin. In the nonrelativistic limit, anyons built from such a formalism reduce to free massive particles in the noncommutative plane \cite{Plyu2}: As in Eq. (\ref{nc}), states with arbitrary spin are indeed associated to noncommutative spatial coordinates. 

\section{Summary and outlook}\label{sec3}

Anyons may exist in $2+1$ dimensions as representations of the Poincar\'e algebra. It follows that they may be present in any Poincar\'e invariant spacetime, including those where a minimal length is present. In such spacetimes, an anyon wave equation can be defined. 

As an outlook, let us mention that previously known analytical results about the spectra of Schr\"odinger-like Hamiltonians with modified Heisenberg algebra could be re-analysed in the particular case of two spatial dimensions to better appraise the interplay between of an arbitrary spin and a minimal length.

An other way of producing anyons is to minimally couple a particle to a vortex-like gauge field \cite{Wilczek}. It could be interesting to see what is the generalization of such a gauge field configuration in electrodynamic with a minimal length, which has been proposed recently in \cite{Qedmin}; we leave such a task for future works.

\acknowledgments
The author thanks N. Boulanger for valuable discussions about anyons and C. Semay for comments about the present manuscript.

\end{document}